\documentstyle[12 pt]{article}
\begin {document}
\title{
Green Function for a Spin $1\over 2$ Particle in a Coulomb + Scalar Potential   }
\author{Arvind Narayan Vaidya$^{\star}$\\
Luiz Eduardo Silva Souza\\
Instituto de F\'\i sica - Universidade Federal do Rio de Janeiro \\
Caixa Postal 68528 - CEP 21945-970, Rio de
Janeiro, Brazil} 
\date{}
\maketitle
\begin{abstract}
We calculate the  Green function for the Dirac equation  describing a spin 1/2 particle in the presence of a potential which is a sum of the Coulomb potential  $ V_C=-A_1/r$ and a Lorentz scalar $V_s=-A_2/r$.
The bound state spectrum is obtained.
   
\end{abstract}
$^{\star}$ {e-mail:vaidya@if.ufrj.br}

\vskip 1.0 cm
\noindent
 {\bf 1. Introduction}
\par
The generalized Dirac equation that we consider describes a spin ${1\over 2}$ particle in interaction with a mixed potential consisting of the Coulomb potential and a Lorentz scalar potential.The scalar potential is added to the mass term in the Dirac equation and may be interpreted as an effective position dependent mass.If the scalar potential is assumed to be created by the exchange of massless scalar mesons, it  has the form $V_S= - {A_2\over r}$.
Exact solutions for the bound states in this mixed potential can be obtained by separation of variables [1,2]. Alternatively, using a similarity transformation it is possible to bring the radial equation to a form almost identical to those of the Schr\"odinger and Klein-Gordon equations in a Coulomb field  leading to the bound state spectrum [3].The bound state problem for the N-dimensional generalized Dirac equation has also been solved [4].The scattering problem does not seem to have been treated in litrature.
\par
In this paper we obtain the explicit form for the Green function  for the generalized Dirac equation.Our results generalize earlier ones for the Dirac-Coulomb Green function [5].
\par
The paper is organized as follows : in Section 2 we formulate the problem,in Section 3 we obtain the projectors which simplify the angular part of the Green function,in Section 4 a Baker-Campbell-Hausdorff formula is used to simplify the radial dependence, in section 5 the final form of the Green function is obtained.Section 6 contains the conclusions.
\vskip 1.0 cm
\noindent

{\bf 2. Formulation of the problem }
\par
 The generalized Dirac equation involves a potential which is  the sum of the time component of a Lorentz vector and a Lorentz scalar and is given by   
\begin{eqnarray}
(\gamma^0(i\partial_0 + {A_1\over r})+i{{\mbox{\boldmath $\gamma$}}\cdot{\mbox{\boldmath $\partial$}}} -M+{A_2\over r})\Psi(x)=0,
\end{eqnarray}
where $\partial_\mu = {\partial\over {\partial x^\mu}}$ and $r=|\bf x|$.
The Green function for this equation satisfies the differential equation 
 \begin{eqnarray}
(\gamma^0(i\partial_0 + {A_1\over r})+i{{\mbox{\boldmath $ \gamma$}}\cdot{\mbox{\boldmath $ \partial$}}} -M+{A_2\over r})G(x,x')= \delta(x-x').
\end{eqnarray}
Since the potential is time independent one may write
\begin{equation}
G(x,x') = {1\over{2\pi}}\int e^{-i\epsilon(t-t')}G({\bf x},{\bf x'}|\epsilon)d\epsilon.
\end{equation}
Thus
\begin{equation}
{G(\bf x,\bf x'|\epsilon)}= ({\gamma^0(\epsilon + {A_1\over r})-M+{A_2\over r}}+i{{\mbox{\boldmath $ \gamma$}}\cdot{\mbox{\boldmath $ \partial$}}} )^{-1}\delta(\bf x-\bf x').
\end{equation}
We rewrite the above equation in the form
\begin{eqnarray}
{G(\bf x,\bf x'|\epsilon)}&=& (\gamma^0(\epsilon + {A_1\over r})+M-{A_2\over r}+i{\mbox{\boldmath $ \gamma$}}\cdot{\mbox{\boldmath $ \partial$}} )\nonumber\\
&\times&( r(\epsilon^2-M^2-{p_r}^2)+2(\epsilon A_1+MA_2)
-{1\over r}{\cal K} )^{-1}\nonumber\\&\times&
{1\over r'}\delta(r-r')\delta ({\bf n}-{\bf n'}),
\end{eqnarray}
where ${\bf n}= {{\bf r}\over r}$, $p_r = -i(\partial_r +{1\over r})$ and
\begin {equation}
{\cal K} = {\bf L^2}-i({A_1}{\gamma^0}+A_2){\mbox{\boldmath $ \gamma$}}\cdot{\mbox{\boldmath $n$}}-({A_1}^2-{A_2}^2).
\end {equation}
\par
Next we use the inegral representation
\begin{equation}
{1\over D} = -i\int_0^{\infty} ds \exp isD ,
\end{equation}
where a convergence factor (not shown explicitly) is assumed to be present in the integral.Then we can write equation (5) as
\begin{eqnarray}
{G(\bf x,\bf x'|\epsilon)}&=& -i (\gamma^0(\epsilon + {A_1\over r})+M-{{A_2\over r}+i{\mbox{\boldmath $ \gamma$}}}\cdot{\mbox{\boldmath $ \partial$}} )\nonumber\\
&\int_0^{\infty}& ds {e^{2is(\epsilon A_1+MA_2)}}{\exp -is( r(k^2+{p_r}^2)
+{1\over r}{\cal K})}
{1\over r'}\delta(r-r')\delta ({\bf n}-{\bf n'}),
\end{eqnarray}
where $k^2=M^2-\epsilon^2$.We assume k to be real.

{\bf 3. The projectors for the eigenvalues of the operator $\cal K$}
\par
In this section we construct the eigenfunctions of the operator $\cal K$ and show that the angular delta function $\delta({\bf n}-{\bf n'})$ may be written as the sum of the projectors of its eigenvalues. 
This may be done easily if we introduce the total angular momentum by $ {\bf J}={\bf L} + {\bf S}={\bf L}+{1\over 2}{\mbox{\boldmath $ \Sigma$}}$ and use a basis in which the operators $\bf J^2,\bf L^2,\bf S^2$ and $J_z$ are diagonal.Using the standard representation for the Dirac matrices and the usual notation for the eigenvalues
of the operators $\bf J^2,\bf L^2$ and $J_z$ the basis functions may be written as 
\begin{eqnarray}
{\Omega_{j,l,m}}^{(+)}&=&\pmatrix {\Omega_{j,l,m}\cr 0},\nonumber\\
{\Omega_{j,l,m}}^{(-)}&=&\pmatrix {0\cr \Omega_{j,l,m}},
\end{eqnarray}
where $\Omega_{j,l,m}$ are the two component spinors 
\begin{equation}
\Omega_{j,l,m} = \pmatrix {{({{l+{1\over 2}+m}\over {2l+1}})^{1\over 2}}Y_{l,{m-{1\over 2}}}\cr {({{l+{1\over 2}-m}\over {2l+1}})^{1\over 2}}Y_{l,{m+{1\over 2}}}},
\end{equation}
for $ j = l+{1\over 2}$,$l=0,1,2,....$ and
\begin{equation}
\Omega_{j,l,m} = \pmatrix {{({{l+{1\over 2}-m}\over {2l+1}})^{1\over 2}}Y_{l,{m-{1\over 2}}}(\theta,\phi)\cr -{({{l+{1\over 2}+m}\over {2l+1}})^{1\over 2}}Y_{l,{m+{1\over 2}}}(\theta,\phi)},
\end{equation}
for $ j = l-{1\over 2}$,$l= 1,2,3,.....$,
which satisfy the equations
\begin{equation}
(1+{{\mbox{\boldmath $ \sigma$}}\cdot{\bf L}}){\Omega_{j,j\pm{1\over 2},m}}= \mp(j +{1\over 2}){\Omega_{j,j\pm{1\over 2},m}},
\end{equation}
\begin{equation}
{{\mbox{\boldmath $ \sigma$}}\cdot{\bf n}}{\Omega_{j,j\pm{1\over 2},m}}= {\Omega_{j,j\mp{1\over 2},m}}.
\end{equation}
In the above basis
\begin{equation}
 {\bf J^2}+{1\over 4}=(1+{{\mbox{\boldmath $ \Sigma$}}\cdot{\bf L}})^2.
\end {equation}
Since the operator $\cal K$ may be written as 
\begin {equation}
{\cal K}= {\bf J^2}+{1\over 4}-(A_1^2-A_2^2)-(1+{{\mbox{\boldmath $ \Sigma$}}\cdot{\bf L}})-i(A_1\gamma^0+A_2){\mbox{\boldmath $ \gamma$}}\cdot{\mbox{\boldmath $n$}},
\end{equation}
we get
\begin{equation}
{\cal K} = \Lambda(\Lambda + 1),
\end{equation}
where 
\begin{equation}
\Lambda = -(1+{{\mbox{\boldmath $ \Sigma$}}\cdot{\bf L}})-i(A_1\gamma^0+A_2){\mbox{\boldmath $ \gamma$}}\cdot{\mbox{\boldmath $n$}}.
\end{equation}
Since 
\begin{equation}
\Lambda^2 = {\bf J^2}+{1\over 4}-(A_1^2-A_2^2),
\end{equation}
we get the eigenvalues of $\Lambda$  as $\pm\lambda$  where 
\begin{equation}
\lambda = [(j+{1\over 2})^2-(A_1^2-A_2^2)]^{1\over 2}.
\end{equation}
We assume that $\lambda$ is real.
\par
Next, one can show that
\begin{equation}
S \Lambda S^{-1}=-(1-{{A_1^2-A_2^2}\over \Gamma^2})^{1\over 2}(1+{{\mbox{\boldmath $ \Sigma$}}\cdot{\bf L}}),
\end{equation}
where
\begin{equation}
S=e^{i{\gamma^0}{{\mbox{\boldmath $ \gamma$}}\cdot{\bf n}}\eta(\Gamma)}e^{i{{\mbox{\boldmath $ \gamma$}}\cdot{\bf n}}\xi(\Gamma)},
\end{equation}
and
\begin{eqnarray}
\Gamma&=&\gamma^0(1+{{\mbox{\boldmath $ \Sigma$}}\cdot{\bf L}}),\nonumber\\
\tanh 2\xi(\Gamma)&=&{A_1\over \Gamma},\nonumber\\
\tan 2\eta(\Gamma)&=&{A_2\over \Gamma (1-{A_1^2\over \Gamma^2})^{1\over 2}}.
\end{eqnarray}
Hence
\begin{eqnarray}
\Gamma{\Omega^{\pm}}_{j,j+{\mu\over 2},m}&=&\mp\mu(j+{1\over 2}){\Omega^{\pm}}_{j,j+{\mu\over 2},m},\nonumber\\
\Lambda S^{-1}{\Omega^{\pm}}_{j,j+{\mu\over 2},m}&=&\mu\lambda S^{-1}{\Omega^{\pm}}_{j,j+{\mu\over 2},m}.
\end{eqnarray}
Thus $S^{-1}{\Omega^{\pm}}_{j,j+{\mu\over 2},m}$ are eigenfunctions of $\Lambda$ with eigenvalues $ \mu \lambda$.
To calculate their explicit form one can use the results
\begin{eqnarray}
{\gamma^0}{\Omega^{\pm}}_{j,j+{\mu\over 2},m}&=&\pm{\Omega^{\pm}}_{j,j+{\mu\over 2},m}\nonumber\\
{{\mbox{\boldmath $ \gamma$}}\cdot{\bf n}}{\Omega^{\pm}}_{j,j+{\mu\over 2},m}&=&\mp {\Omega^{\mp}}_{j,j-{\mu\over 2},m}\nonumber\\
{\gamma^0}{{\mbox{\boldmath $ \gamma$}}\cdot{\bf n}}{\Omega^{\pm}}_{j,j+{\mu\over 2},m}&=& {\Omega^{\mp}}_{j,j-{\mu\over 2},m}
\end{eqnarray}
Using  equations (21-23) the unnormalized  eigenfunctions of $\Lambda$ belonging to the eigenvalues $\pm\lambda$ are
\begin{eqnarray}
{\Phi_{\lambda,j,m}}^{(\pm)} &=& {{\lambda+j+{1\over 2}}\over 2\lambda }{\Omega_{j,{j+{1\over 2}},m}}^{(\pm)} -i{{A_1 \mp A_2}\over 2\lambda}
{\Omega_{j,{j-{1\over 2}},m}}^{(\mp)}\nonumber\\
{\Phi_{{-\lambda},j,m}}^{(\pm)} &=& {{\lambda+j+{1\over 2}}\over 2\lambda }{\Omega_{j,{j-{1\over 2}},m}}^{(\pm)} +i{{A_1 \mp A_2}\over 2\lambda}
 {\Omega_{j,{j+{1\over 2}},m}}^{(\mp)}
\end{eqnarray}
Next we have the completeness relation 
\begin{equation}
{\delta ( \bf n-\bf n')} = \Sigma_{j,\mu,m}{\Omega_{j,j+{\mu\over 2},m}}^{(+)}{(\bf n)}{{\Omega_{j,j+{\mu\over 2},m}}^{(+)}{(\bf n')}}\dagger + {\Omega_{j,j+{\mu\over 2},m}}^{(-)}{(\bf n)}{{\Omega_{j,j+{\mu\over 2},m}}^{(-)}{(\bf n')}}\dagger
\end{equation}
where $\mu = \pm 1$.
Next, inverting  equation (25) we have
\begin{eqnarray}
{\Omega_{j,j+{\mu\over 2},m}}^{(+)}&=&{\Phi_{\mu\lambda,j,m}}^{(+)}+ i\mu {{A_1-A_2}\over {\lambda+j+{1\over 2}}} {\Phi_{-\mu\lambda,j,m}}^{(-)}\nonumber\\
{\Omega_{j,j+{\mu\over 2},m}}^{(-)}&=&{\Phi_{\mu\lambda,j,m}}^{(-)}+ i\mu {{A_1+A_2}\over {\lambda+j+{1\over 2}}}{\Phi_{-\mu\lambda,j,m}}^{(+)}
\end{eqnarray}
Hence 
\begin{eqnarray}
\Lambda P_{\pm\lambda}&=&\pm\lambda P_{\pm\lambda}\nonumber\\
{\delta ( \bf n-\bf n')}&=& \Sigma_{j}(P_{\lambda}{(\bf n,\bf n')} +\ P_{-\lambda}{(\bf n,\bf n')})
\end{eqnarray}
where the projectors $P_{\pm \lambda}$ are given by
\begin{eqnarray}
P_{\lambda}({\bf n},{\bf n'})&=&{\Sigma_m} {\Phi_{\lambda,j,m}}^{(+)}({\bf n})
( {\Omega_{j,j+{1\over 2},m}}^{(+)}({\bf n'})\dagger-i{{A_1+A_2}\over {\lambda+j+{1\over 2}}}{\Omega_{j,j-{1\over 2},m}}^{(-)}(\bf n')\dagger)\nonumber\\
&+& {\Phi_{\lambda,j,m}}^{(-)}({\bf n})
( {\Omega_{j,j+{1\over 2},m}}^{(-)}({\bf n'})\dagger-i{{A_1-A_2}\over {\lambda+j+{1\over 2}}}{\Omega_{j,j-{1\over 2},m}}^{(+)}(\bf n')\dagger)\nonumber\\
P_{-\lambda}({\bf n},{\bf n'})&=& {\Sigma_m}{\Phi_{-\lambda,j,m}}^{(+)}({\bf n})
( {\Omega_{j,j-{1\over 2},m}}^{(+)}({\bf n'})\dagger+i{{A_1+A_2}\over {\lambda+j+{1\over 2}}}{\Omega_{j,j+{1\over 2},m}}^{(-)}(\bf n')\dagger)\nonumber\\
&+& {\Phi_{-\lambda,j,m}}^{(-)}({\bf n})
( {\Omega_{j,j-{1\over 2},m}}^{(-)}({\bf n'})\dagger+i{{A_1-A_2}\over {\lambda+j+{1\over 2}}}{\Omega_{j,j+{1\over 2},m}}^{(+)}(\bf n')\dagger)
\end{eqnarray}
 If we now use equation (25) one can write the projectors in terms of the $\Omega$ functions.
The summations involved  may be done as follows.First using equations (11) and (12) we get
\begin{eqnarray}
{4\pi}{\Sigma_m}\Omega_{j,j+{\mu\over 2},m}{(\bf n)}\Omega_{j,j+{\mu\over 2},m}{(\bf n')}\dagger = (j+{1\over 2})P_{j+{\mu\over 2}}{(x)} +i\mu {\mbox{\boldmath $ \sigma$}}\cdot{(\bf n\times\bf n')}P_{j+{\mu\over 2}}^{'}{(x)}
\end{eqnarray}
where $x = {\bf n\cdot \bf n'}$ and the prime on the Legendre polynomials $P_{j+{\mu\over 2}}$  denotes derivative with respect to x.
Also,
\begin{eqnarray}
(j+{1\over 2})P_{j+{1\over 2}}(x) &=& x{P_{j+{1\over 2}}}^{'}(x) - P_{j-{1\over 2}}^{'}(x)\nonumber\\
(j+{1\over 2})P_{j-{1\over 2}}(x) &=& {P_{j+{1\over 2}}}^{'}(x) - x{P_{j-{1\over 2}}}^{'}(x)
\end{eqnarray}
Hence
 \begin{eqnarray}
{4\pi}{\Sigma_{\mu,m}}\mu\Omega_{j,j+{\mu\over 2},m}{(\bf n)}\Omega_{j,j+{\mu\over 2},m}{(\bf n')}\dagger = (-1 + x +i {\mbox{\boldmath $ \sigma$}}\cdot{(\bf n\times\bf n')})A_j
\end{eqnarray}
\begin{eqnarray}
{4\pi}{\Sigma_{\mu,m}}\Omega_{j,j+{\mu\over 2},m}{(\bf n)}\Omega_{j,j+{\mu\over 2},m}{(\bf n')}\dagger = (1 + x +i {\mbox{\boldmath $ \sigma$}}\cdot{(\bf n\times\bf n')}) B_j
\end{eqnarray}
where
\begin{eqnarray}
A_j &=&{j+{1\over 2}\over x-1} (P_{j+{1\over 2}}(x) -P_{j-{1\over 2}}(x))\nonumber\\
B_j &=&{j+{1\over 2}\over x+1} (P_{j+{1\over 2}}(x)+P_{j-{1\over 2}}(x))
\end{eqnarray}
The final expression for $P_{\lambda}$ is
\begin{eqnarray}
8\pi P_{\lambda} &=&(1 + x +i {\mbox{\boldmath $ \Sigma$}}\cdot{(\bf n\times\bf n')})B_j
+{{j+{1\over 2}}\over \lambda}(-1 + x +i {\mbox{\boldmath $ \Sigma$}}\cdot{(\bf n\times\bf n')}) A_j\nonumber\\
&-&i{(A_1\gamma^0+A_2)\over \lambda}{\mbox{\boldmath $ \gamma$}}\cdot{(\bf n+\bf n')}B_j
\end{eqnarray}
The results for the projector $P_{-\lambda}$ can be obtained from the above equation by changing the sign of $\lambda$. 
 The projectors $P_{\pm\lambda}$ satisfy the relations:
\begin{eqnarray}
\int d{\bf n'} P_{\pm\lambda}({\bf n},{\bf n'})P_{\pm\lambda'}({\bf n'},{\bf n''})&=& \delta_{\lambda\lambda'}P_{\pm\lambda}({\bf n},{\bf n''})\nonumber\\
\int d{\bf n'} P_{\lambda}({\bf n},{\bf n'})P_{-\lambda'}({\bf n'},{\bf n''})&=& 0
\end{eqnarray}
Thus equation (8) can now be written in the form
\begin{eqnarray}
{G(\bf x,\bf x'|\epsilon)}&=& -i (\gamma^0(\epsilon + {A_1\over r})+M-{{A_2\over r}+i{\mbox{\boldmath $ \gamma$}}}\cdot{\mbox{\boldmath $ \partial$}} )\Sigma_{\nu,j}P_{\nu\lambda}{(\bf n,\bf n')}
\nonumber\\
&\int_0^{\infty}& ds {e^{2is(\epsilon A_1+MA_2)}}{\exp -is( r(k^2+{p_r}^2)
+\nu\lambda(\nu\lambda +1){1\over r})}
{1\over r'}\delta(r-r')
\end{eqnarray}
where $\nu = \pm 1$
\vskip 1.0 cm
\noindent
{\bf 4 The Baker-Campbell-Hausdorff formula for the SO(2,1) algebra}
\par
In this section we calculate the action of the operator $
\exp -is( r{p_r}^2
+\nu\lambda(\nu\lambda +1){1\over r})$ on a suitable function $f(r)$.This will allow us to simplify the radial dependence of the Green function.The method used is the
same as one used by Mil'shtein and Strakhovenko previously.We include the calculation for the sake of completeness. 
The operators 
\begin{eqnarray}
T_1 &=& {1\over 2}(r{p_r}^2
+\nu\lambda(\nu\lambda +1){1\over r})\nonumber\\
T_2 &=&r{p_r}\nonumber\\
T_3 &=& r
\end{eqnarray}
obey the commutation rules
\begin{eqnarray}
\lbrack T_1,T_2 \rbrack &=& -iT_1\nonumber\\
\lbrack T_1,T_3 \rbrack &=& -iT_2\nonumber\\
\lbrack T_2,T_3 \rbrack &=& -iT_3
\end{eqnarray}
which are equivalent to those of the generators of the $SO(2,1)$ algebra.
\par
In the notation introduced above we need to calculate the action of the operator $\exp -2is(T_1 + {k^2\over 2}T_3)$ on a function $f(r)$.
Next we have 
\begin{equation}
\exp -2is(T_1 + {k^2\over 2}T_3) = e^{-iaT_3}e^{-ibT_2}e^{-icT_1}
\end{equation}
where
\begin{eqnarray}
a&=& k \tan ks\nonumber\\
b&=& 2ln \cos ks \nonumber\\
c&=& ({2\over k})\cot ks
\end{eqnarray}
as may be easily verified using the $2\times 2$ matrix representation of the operators $ T_1,T_2 $ and $T_3$.
Since
\begin{equation}
e^{-ibT_2} \phi(r) = e^{-b} \phi(re^{-b})
\end{equation}
we need only to calculate explicitly the action of the operator $ e^{-icT_1}$ on a function $f(r)$
Assume that $ f(r)$ admits the Laplace transform
\begin{eqnarray}
f(r) &=& {1\over {2i\pi}}\int_{-i{\infty}}^{i{\infty}}d\sigma F(\sigma)e^{\sigma r} r^{\delta}\nonumber\\
F(\sigma)&=& \int_0^{\infty}dr f(r) e^{-\sigma r }r^{-\delta}
\end{eqnarray}
where
\begin{equation}
T_1 r^\delta = 0
\end{equation}.
For $\nu = 1$ we choose $\delta = \lambda $ and for $\nu = -1$ we take $ \delta = \lambda - 1 $.
\par 
Next,
\begin{equation}
e^{-icT_1}e^{\sigma T_3} = e^{-ia_1T_3} e^{-ib_1T_2} e^{-ic_1T_1}
\end{equation}
where
\begin{eqnarray}
a_1&=& {i\sigma\over{1-{1\over 2}i\sigma c}}\nonumber\\
b_1&=& 2 ln (1-{1\over 2}i\sigma c )\nonumber\\
c_1&=& {c\over (1-{1\over 2}i\sigma c)}
\end{eqnarray}
as may be easily verified using the $2\times 2$ matrix representation of the operators $ T_1,T_2 $ and $T_3$.
Thus
\begin{equation}
 e^{-icT_1}e^{\sigma T_3}r^\delta = {r^\delta\over (1-{1\over 2}i\sigma c)^{2(\delta + 1)}}\exp {{\sigma r \over (1-{1\over 2}i\sigma c)}}
\end{equation}
Hence
\begin{equation}
{e^{-iuT_1}}f(r) = \int_0^{\infty}dr' f(r') e^{-\sigma r'+ {2ir\over c} }
{{({r\over r'})}^\delta}{1\over {2i\pi}}\int_{-i{\infty}}^{i{\infty}}d\sigma{1\over (1-{1\over 2}i\sigma c)^{2(\delta + 1)}}
{\exp {-2ir \over c(1-{1\over 2}i\sigma c)}}
\end{equation}
Expanding the exponential on the right hand side in series we  get a series of integrals which can be evaluated by the residue theorem so that we finally get 
\begin{equation}
{e^{-icT_1}}f(r) = -{2\over c}\int_0^{\infty}dx f(x) {({x\over r})^{1\over 2}}{J_{2\delta + 1}({4\over c}(xr)^{1\over 2}){e^{{2i\over c}(x + r)}e^{-i\pi\delta}}}
\end{equation}
We now choose $f(r) = \delta (r- r')$ which gives 
\begin{equation}
 \exp -2is(T_1 + {k^2\over 2}T_3)\delta(r-r') = -{k\over \sin ks}({r'\over r})^{1\over 2}J_{2\delta + 1}({2k\over \sin ks}(rr')^{1\over 2})e^{ik\cot ks(r + r')}e^{-i\pi\delta}
\end{equation}
Thus equation (37) can now be written in the form
\begin{eqnarray}
{G(\bf x,\bf x'|\epsilon)}&=& i (\gamma^0(\epsilon + {A_1\over r})+M-{{A_2\over r}+i{\mbox{\boldmath $ \gamma$}}}\cdot{\mbox{\boldmath $ \partial$}} )
\int_0^{\infty} ds {e^{2is(\epsilon A_1+mA_2)}}e^{ik\cot ks(r + r')}e^{-i\pi\lambda}\nonumber\\
&\times&{k\over \sin ks}({1\over rr'})^{1\over 2}
\Sigma_{\nu,j}{\nu }P_{\nu\lambda}{(\bf n,\bf n')}J_{2\lambda+\nu}({2k\over \sin ks}(rr')^{1\over 2})
\end{eqnarray}
where $\nu = \pm 1$
It remains now to work out the action of the differential operators.
\vskip 1.0 cm
\noindent
{\bf 5.The final form of the Green function} 
\par
To simplify the Green function we note that
\begin{equation}
-i{\mbox{\boldmath $ \gamma$}}\cdot{\mbox{\boldmath $ \partial$}}= (p_r - {i\over r}(1+{{\mbox{\boldmath $ \Sigma$}}\cdot{\bf L}})){\mbox{\boldmath $ \gamma$}}\cdot{\mbox{\boldmath $ n$}}={\mbox{\boldmath $ \gamma$}}\cdot{\mbox{\boldmath $ n$}}(p_r + {i\over r}(1+{{\mbox{\boldmath $ \Sigma$}}\cdot{\bf L}}))
\end{equation}
Using equation (17) we get
\begin{eqnarray}
-i{\mbox{\boldmath $ \gamma$}}\cdot{\mbox{\boldmath $ \partial$}}= {\mbox{\boldmath $ \gamma$}}\cdot{\mbox{\boldmath $ n$}}(p_r - {i\over r}\Lambda) +{1\over r}(A_1\gamma^0-A_2)
\end{eqnarray}
Hence
\begin{eqnarray}
(\gamma^0(\epsilon + {A_1\over r})+M-{{A_2\over r}+i{\mbox{\boldmath $ \gamma$}}}\cdot{\mbox{\boldmath $ \partial$}} )=\gamma^0\epsilon+M-{\mbox{\boldmath $ \gamma$}}\cdot{\mbox{\boldmath $ n$}}(p_r - {i\over r}\Lambda)
\end{eqnarray}
Using equation (23) we can prove the following relations:
\begin{eqnarray}
{8\pi}{\mbox{\boldmath $ \gamma$}}\cdot{\bf n}P_{\pm\lambda} &=& B_j{\mbox{\boldmath $ \gamma$}}\cdot{(\bf n+\bf n')} \mp{{j+{1\over 2}}\over \lambda}A_j{\mbox{\boldmath $ \gamma$}}\cdot{(\bf n-\bf n')}\nonumber\\
&\mp&{i\over \lambda}(A_1\gamma^0-A_2)(1 + x +i{\mbox{\boldmath $\Sigma$}}\cdot{(\bf n\times\bf n')})B_j
\end{eqnarray}
\begin{eqnarray}
\lbrack (rr')^{1\over 2}e^{-ik\cot ks(r+r')}\rbrack
(p_r\mp{i\lambda\over r}) \lbrack (rr')^{-1\over 2}J_{2\lambda\pm 1}(y)e^{ik\cot ks(r+r')}\rbrack\nonumber\\
 =
k\cot ks J_{2\lambda\pm 1}(y)\mp{iy\over 2r}{{J}_{2\lambda}}(y)
\end{eqnarray}
where $y = {{2k(rr')^{1\over 2}}\over \sin ks}$.
 Hence
\begin{eqnarray}
 &(rr')^{1\over 2}&e^{-ik\cot ks(r+r')}
({\gamma^0}(\epsilon+{A_1\over r}) +M-{A_2\over r}+i{\mbox{\boldmath $ \gamma$}}\cdot{\mbox{\boldmath $ \partial$}} )
 (rr')^{-1\over 2}e^{ik\cot ks(r+r')}{8\pi}(J_{2\lambda+1}(y){P_{\lambda}(\bf n,\bf n')}=\nonumber\\
 &(&\gamma^0\epsilon +M)J_{2\lambda+1}(y)\nonumber\\
&\lbrack&(1 + x +i {\mbox{\boldmath $ \Sigma$}}\cdot{(\bf n\times\bf n')})B_j
+{{j+{1\over 2}}\over \lambda}(-1 + x +i {\mbox{\boldmath $ \Sigma$}}\cdot{(\bf n\times\bf n')}) A_j
-i{(A_1\gamma^0+A_2)\over \lambda}{\mbox{\boldmath $ \gamma$}}\cdot{(\bf n+\bf n')}B_j)\rbrack\nonumber\\
&-&(k\cot ks J_{2\lambda+1}(y)-{iy\over 2r}{J_{2\lambda}}(y))\nonumber\\
&\times&\lbrack B_j{\mbox{\boldmath $ \gamma$}}\cdot{(\bf n+\bf n')}-{(j+{1\over 2})\over \lambda}A_j{\mbox{\boldmath $ \gamma$}}\cdot{(\bf n-\bf n')}
-{i\over \lambda}(A_1\gamma^0-A_2)(1 + x +i{\mbox{\boldmath $\Sigma$}}\cdot{(\bf n\times\bf n')})B_j\rbrack
\end{eqnarray}
and
\begin{eqnarray}
 &(rr')^{1\over 2}&e^{-ik\cot ks(r+r')}
({\gamma^0}(\epsilon+{A_1\over r}) +M-{A_2\over r}+i{\mbox{\boldmath $ \gamma$}}\cdot{\mbox{\boldmath $ \partial$}} )
 (rr')^{-1\over 2}e^{ik\cot ks(r+r')}{8\pi}(J_{2\lambda-1}(y){P_{-\lambda}(\bf n,\bf n')}=\nonumber\\
 &(&\gamma^0\epsilon +M)J_{2\lambda-1}(y)\nonumber\\
&\lbrack&(1 + x +i {\mbox{\boldmath $ \Sigma$}}\cdot{(\bf n\times\bf n')})B_j
-{{j+{1\over 2}}\over \lambda}(-1 + x +i {\mbox{\boldmath $ \Sigma$}}\cdot{(\bf n\times\bf n')}) A_j
+i{(A_1\gamma^0+A_2)\over \lambda}{\mbox{\boldmath $ \gamma$}}\cdot{(\bf n+\bf n')}B_j)\rbrack\nonumber\\
&-&(k\cot ks J_{2\lambda-1}(y)+{iy\over 2r}{J_{2\lambda}}(y))\nonumber\\
&\times&\lbrack B_j{\mbox{\boldmath $ \gamma$}}\cdot{(\bf n+\bf n')}+{(j+{1\over 2})\over \lambda}A_j{\mbox{\boldmath $ \gamma$}}\cdot{(\bf n-\bf n')}
+{i\over \lambda}(A_1\gamma^0-A_2)(1 + x +i{\mbox{\boldmath $\Sigma$}}\cdot{(\bf n\times\bf n')})B_j\rbrack
\end{eqnarray}
Hence we get
\begin{eqnarray}
{G(\bf x,\bf x'|\epsilon)}&=& {-i\over 4\pi rr'}\Sigma_j  
\int_0^{\infty} ds {e^{2is(\epsilon A_1+mA_2)}}e^{ik\cot ks(r + r')}e^{-i\pi\lambda}  T_j(r,r',s)
\end{eqnarray}
where
\begin{eqnarray}
&{{-4}\over y}& T_j(r,r',s)=\nonumber\\
&(&\gamma^0\epsilon +M)J_{2\lambda+1}(y)\nonumber\\
&\lbrack&(1 + x +i {\mbox{\boldmath $ \Sigma$}}\cdot{(\bf n\times\bf n')})B_j
+{{j+{1\over 2}}\over \lambda}(-1 + x +i {\mbox{\boldmath $ \Sigma$}}\cdot{(\bf n\times\bf n')}) A_j
-i{(A_1\gamma^0+A_2)\over \lambda}{\mbox{\boldmath $ \gamma$}}\cdot{(\bf n+\bf n')}B_j)\rbrack\nonumber\\
&-&(k\cot ks J_{2\lambda+1}(y)-{iy\over 2r}{J_{2\lambda}}(y))\nonumber\\
&\times&\lbrack B_j{\mbox{\boldmath $ \gamma$}}\cdot{(\bf n+\bf n')}-{(j+{1\over 2})\over \lambda}A_j{\mbox{\boldmath $ \gamma$}}\cdot{(\bf n-\bf n')}
-{i\over \lambda}(A_1\gamma^0-A_2)(1 + x +i{\mbox{\boldmath $\Sigma$}}\cdot{(\bf n\times\bf n')})B_j\rbrack\nonumber\\
&-(&\gamma^0\epsilon +M)J_{2\lambda-1}(y)\nonumber\\
&\lbrack&(1 + x +i {\mbox{\boldmath $ \Sigma$}}\cdot{(\bf n\times\bf n')})B_j
-{{j+{1\over 2}}\over \lambda}(-1 + x +i {\mbox{\boldmath $ \Sigma$}}\cdot{(\bf n\times\bf n')}) A_j
+i{(A_1\gamma^0+A_2)\over \lambda}{\mbox{\boldmath $ \gamma$}}\cdot{(\bf n+\bf n')}B_j)\rbrack\nonumber\\
&+&(k\cot ks J_{2\lambda-1}(y)+{iy\over 2r}{J_{2\lambda}}(y))\nonumber\\
&\times&\lbrack B_j{\mbox{\boldmath $ \gamma$}}\cdot{(\bf n+\bf n')}+{(j+{1\over 2})\over \lambda}A_j{\mbox{\boldmath $ \gamma$}}\cdot{(\bf n-\bf n')}
+{i\over \lambda}(A_1\gamma^0-A_2)(1 + x +i{\mbox{\boldmath $\Sigma$}}\cdot{(\bf n\times\bf n')})B_j\rbrack
\end{eqnarray}
One may rewrite the above result in the form
\begin{eqnarray}
T_j(r,r',s)&=&
(1 + x +i {\mbox{\boldmath $ \Sigma$}}\cdot{(\bf n\times\bf n')})B_j({1\over 2}({\gamma^0}\epsilon +M)y J'_{2\lambda}(y)-i(A_1\gamma^0-A_2)k\cot ksJ_{2\lambda})\nonumber\\
&+&(j+{1\over 2})(1 - x -i {\mbox{\boldmath $ \Sigma$}}\cdot{(\bf n\times\bf n')}) A_j J_{2\lambda}({\gamma^0}\epsilon+M)\nonumber\\
&+&i({\gamma^0}\epsilon + M)(A_1\gamma^0+A_2){\mbox{\boldmath $ \gamma$}}\cdot{(\bf n+\bf n')}B_jJ_{2\lambda}\nonumber\\
&-&k\cot ks({\mbox{\boldmath $ \gamma$}}\cdot{(\bf n+\bf n')}B_j{1\over 2}yJ'_{2\lambda}-(j+{1\over 2})A_j{\mbox{\boldmath $ \gamma$}}\cdot{(\bf n-\bf n')}J_{2\lambda})\nonumber\\
&-&{{iy^2}\over 4r} J_{2\lambda}{\mbox{\boldmath $ \gamma$}}\cdot{(\bf n+\bf n')}B_j
\end{eqnarray}
where the prime on the Bessel functions $J_{2\lambda}(y)$ means derivative with respect to $y$.
\par 
The integral over $s$ can be decomposed into the sum of integrals over the segments of length ${\pi\over k}$.The Bessel function passes from sheet to sheet at points $s= n{\pi\over k}$ at integer $n=1,2,...$ acquiring a phase $e^{-2i\pi\lambda}$.Using this fact and the periodicity of $\cot ks$ the above expression for $G(\bf r,\bf r'|\epsilon)$ can be transformed to the form 
\begin{eqnarray}
{G(\bf r,\bf r'|\epsilon)}&=& {1\over 8\pi r r' k}\Sigma_j{1\over \sin {\pi({{\epsilon A_1+MA_2}\over k}-\lambda)}}\nonumber\\
&\times&{\int_{-{\pi\over 2}}^{\pi\over 2}} d\tau 
{e^{ik(r+r')\cot \tau-2i\tau{{\epsilon A_1+MA_2}\over k}}}{T_j(r,r',{(\pi-2\tau)\over  2k})}
\end{eqnarray}
The Green function obtained above has simple poles at the points 
\begin{eqnarray}
{\epsilon_{ n}\over M}=-{A_1A_2\over ({\hat n}^2+{A_1}^2)}\pm\lbrack({A_1A_2\over {\hat n}^2+{A_1}^2})^2+{{\hat n }^2-A_2^2\over {\hat n}^2+{A_1}^2}\rbrack^{1\over 2}
\end{eqnarray}
where $\hat n = n+\lambda $ and $n= 0,1,2, .....$
which correspond to bound states.
\vskip 1.0 cm
\noindent
{\bf 6. Conclusion }
\par
In conclusion, we have constructed the Green function for the  Dirac equation for a spin ${1\over 2}$ particle in a mixed potential which is the sum of the Coulomb potential $V_C=-{A_1\over r}$ and a scalar potential  $V_S=-{A_2\over r}$ and obtained the bound state spectrum.The results for the Dirac-Coulomb problem [5] can be reproduced when we put $A_2=0$.  


\end{document}